# Anti-seizure medication load is not correlated with early termination of seizure spread

Nathan Evans[1*], Sarah J. Gascoigne[1], Guillermo M. Besné[1], Chris Thornton[1], Gabrielle M. Schroeder[1], Fahmida A Chowdhury[3], Beate Diehl[3], John S Duncan[3], Andrew W McEvoy[3], Anna Miserocchi[3], Rhys Thomas[2], Jane de Tisi[3], Peter N. Taylor[1,2,3], Yujiang Wang[1,2,3*]

1. CNNP Lab (www.cnnp-lab.com), Interdisciplinary Computing and Complex BioSystems Group, School of Computing, Newcastle University, Newcastle upon Tyne, United Kingdom

2. Faculty of Medical Sciences, Newcastle University, Newcastle upon Tyne, United Kingdom

3. UCL Queen Square Institute of Neurology, Queen Square, London, United Kingdom

* n.evans5@ncl.ac.uk or yujiang.wang@ncl.ac.uk

# Abstract

**Objective:** Anti-seizure medications (ASMs) are the mainstay of treatment for epilepsy, yet their effect on seizure spread is not fully understood. Higher ASM doses have been associated with shorter and less severe seizures. We aimed to test if this effect was due to limiting seizure spread through early termination of otherwise unchanged seizures.

**Methods:** We retrospectively examined intracranial EEG (iEEG) recordings in 15 subjects who underwent ASM tapering during pre-surgical monitoring. We estimated ASM plasma concentrations based on pharmaco-kinetic modeling. In each subject, we identified seizures that followed the same onset and initial spread patterns, but some seizures terminated early (truncated seizures), and other seizures continued to spread (continuing seizures).

We first compared seizure duration to ASM concentration for all seizures and the subset of seizures included in truncated-continuing pairs. Then we compared durations of the matched truncated and continuing seizures. Finally, we compared ASM concentrations at the times of truncated seizures and continuing seizures.

**Results**: Seizure durations were found to be significantly longer at lower ASM concentrations. Continuing seizures were significantly longer in duration than matched truncated seizures. We found no substantial difference between ASM concentrations when truncated *vs.* continuing seizures occurred.

**Significance**: The lack of difference between ASM concentrations at the time of truncated *vs.* continuing seizures implies a separate mechanism for shortening the duration of seizures beyond stopping the spread pathways early. Additionally, the mechanism causing seizures to be truncated remains unclear. Further research is needed to understand how ASM may modulate seizure duration and severity.

**Plain Language Summary:** Anti-seizure medications (ASMs) are taken by people with drug-resistant epilepsy to reduce the rate of seizures. It has been shown before, and reproduced here, that higher ASM doses are correlated with shorter seizures. Additionally, when considering how seizures spread across the brain, we have found that some seizures follow the same spread patterns before a subset stops early. These truncated seizures tend to be shorter. We hypothesised that they would occur at times of higher ASM doses. We found that this was not the case, implying that higher ASM doses are not causing shorter seizures through truncation.

# Key points
- We introduce a system for defining truncated-continuing pairs: seizures sharing the same spread patterns but where one stops early.
- We then used pharmaco-kinetic modeling to compare the ASM concentrations between truncated and continuing seizures.
- There was no significant difference between ASM concentration at the time of truncated vs continuing seizures

- This implies the mechanism by which higher ASM doses shorten seizure durations is not by promoting truncated seizures.

# 1. Introduction

Anti-seizure medications (ASMs) are the most common treatment for epilepsy; however, the effect of ASMs on seizure onset and spread patterns is not fully understood. Expanding on this knowledge could be vital for a range of future treatment approaches such as chronotherapy[1] where medication doses are altered based on the risk of seizures at different times, as well as neurostimulation techniques such as RNS where the seizure onset zone can be physically targeted[2].

Analysis of ictal intracranial EEG (iEEG) has shown an individual-level tendency for seizures to follow specific spread patterns[3,4]. However not every seizure will spread in the same way; instead, there are variations in the pattern[5]. Further, some subjects have truncated seizures that share the same onset and initial spread as other continuing seizures, but terminate early[6,7]. It was our objective to test if ASMs limit seizure spread through this early termination of otherwise unchanged seizures.

Tapering ASMs during iEEG implantation is a common approach to provoke seizures during presurgical monitoring to help identify seizure onset regions[8]. While medication tapering is performed for clinical and monitoring purposes, it can also provide unique insights into seizure dynamics[9]. Higher ASM doses make seizures less likely to occur, and to have a shorter duration when they do occur[10,11]. As truncated seizures tend to have a shorter duration than continuing seizures[7], we hypothesised that truncated seizures are more likely to occur in the presence of higher concentrations of ASMs, and that continuing seizures are associated with lower concentrations. To test this hypothesis, we retrospectively analysed data collected from individuals who were undergoing the medication tapering procedure during iEEG monitoring.

# 2. Methods

## 2.1 Study design

We retrospectively analysed iEEG data from subjects undergoing presurgical evaluation for drug resistant focal epilepsy at the National Hospital for Neurology and Neuroscience. Data were collected for clinical purposes and independent of this study. Our retrospective and anonymised data were obtained from the National Epilepsy & Neurology Database[12], composed of individuals with focal epilepsy from the National Hospital for Neurology and Neurosurgery, and analysed with the approval of Newcastle University Ethics Committee (27742/2022).

Subjects were included if they had undergone iEEG, had subsequent epilepsy surgery, associated pre and post-operative T1w MRI, and iEEG implantation information between 2008 and 2019. Subjects were additionally required to have had ASM tapering with available data on the time and dose of each medication throughout their time in the Epilepsy Monitoring Unit (EMU). We also required that subjects had at least two visually confirmed seizures that had activity captured by our automatic detection algorithms, discussed in Section 2.4. Finally, we matched pairs of truncated and continuing seizures within subjects, and only those subjects with at least one pair of seizures were retained for our analysis. Fifteen subjects met all inclusion criteria.

## 2.2 Subject metadata

Table 1 dispalys demographic information for subjects in this study.

|  | All | Tapered | Tapered with matched pairs |
|---|---|---|---|
| Count | 69 | 42 | 15 |
| Age | 32(15.4) | 31(14.0) | 32(13.8) |
| Sex (M/F) | 33/36 | 20/22 | 7/8 |
| Disease duration | 20(15.0) | 17.5(14.0) | 20(13.5) |
| TLE/eTLE | 31/38 | 27/15 | 9/6 |

***Table 1: Metadata of subjects.*** *Clinical metadata and demographics split by whether ASM tapering was performed during epilepsy monitoring. Data is presented in the following format:*

- ***Count:*** *number of subjects*
- ***Age (years):*** *subject age at time of iEEG exam given as median(interquartile range)*
- ***Sex:*** *given as male/female*
- ***Disease duration (years):*** *Time between epilepsy diagnosis and iEEG exam in years given as median(interquartile range)*
- ***Diagnosis:*** *Temporal Lobe Epilepsy/Extra Temporal Lobe Epilepsy*

## 2.3 Anti-seizure medication tapering and blood plasma concentration estimation

For each subject that underwent ASM tapering, we modelled their ASM plasma concentrations to estimate the relative ASM levels at the time of each seizure.

Each medication dose taken by each subject was recorded during the initial monitoring. A continuous estimation of the plasma concentration for each medication was generated from the intake information using known pharmaco-kinetics. We modeled first order absorption and elimination using equation 1. The parameters used were specific to each medication, and the justification for each parameter is provided in Besne et al.[11].

$$C = \frac{F * D * Ka}{Vd * (Ka - Ke)} * \left(e^{-Ke*t} - e^{-Ka*t}\right) \quad (1)$$

***Equation 1:*** *Pharmaco-kinetic equation for plasma concentration of a single dose oral intake with first-order absorption and elimination. C, plasma concentration. F, bio-availability. D,* *dose at intake. Ka, absorption constant. Ke, elimination constant. t, time since intake. Vd, volume distribution.*

We applied a 6 hour rolling average to computed plasma concentrations to account for inaccuracies in recorded intake time. For each ASM, the level was normalised so that the variation in the pre-tapering regime oscillated between -1 and +1. A single combined ASM load was calculated by summing the different ASMs for each time point, then re-normalised so the combined load ranged between -1 and +1 in the pre-tapering period. This normalised ASM level will be used throughout the rest of this paper.

## 2.4 iEEG preprocessing and ictal activity detection

The iEEG was recorded as part of the presurgical workup. Seizure timings were extracted from clinical reports and visually confirmed. For each seizure, iEEG segments containing ictal and 120 seconds of pre-ictal activity were extracted. Prior to analysis, all data were resampled to 512 Hz. An iterative noise detection algorithm was used to identify pre-ictal noise, which was confirmed by visual inspection. On a within-subject basis, channels identified as noisy were removed from all seizures. When seizures occurred in close succession (seizure cluster), we retained only the lead seizure. Recordings were re-referenced to a common average reference, notch filtered at 50 Hz (and harmonics up to 200 Hz) with a 2 Hz window to remove line noise. Recordings were then band-pass filtered between 0.5 and 200 Hz. All filters used a fourth-order, zero-phase shift Butterworth filter.

Our method for detecting ictal activity was an extension of the previously-described seizure "imprint" algorithm[13]. Briefly, we first computed eight EEG signal features: energy, line length, and band power in six distinct bands ($\delta, \theta, \alpha, \beta$, low-$\gamma$, and high-$\gamma$) in one second windows with a 7/8th second overlap. The distribution in all features across all pre-ictal windows was then used to calculate the level of abnormality for each window in the ictal segments. By thresholding the abnormality, this approach produces a binary "imprint" of the seizure showing where seizure activity (defined as abnormal activity compared to pre-ictal) has been detected at each time point in each channel (Fig. 1A). The details of the algorithm are documented in code form for one example seizure on Github:

https://github.com/NathanMEvans/Temporal-Modulation-truncation/blob/main/imprint/generate_imprints_SWEZ.m.

## 2.5 Matching early terminating seizures with continuing seizures

We matched truncated seizures with continuing seizures that shared the same onset and initial propagation patterns in their Imprint (Fig. 1A for an example pair of seizures). As seizures do not always propagate at the same rate[4,7,14], we used dynamic time warping (DTW) to compare Imprints. DTW allows for the comparison of time series without penalising one series progressing faster or slower than the other. To allow partial comparisons, i.e. comparing the entirety of one seizure to the beginning of another, we used open-ended DTW (OE-DTW)[15]. This was done as a truncated seizure would theoretically have an Imprint that matched just the start of a continuing seizure that shared the same pathway. Fig. 1B is an example of how DTW finds the minimum cumulative distance between the two Imprints (with the matching between time points shown as a blue solid line). Comparatively the OE-DTW matching is shown in red, only comparing the beginning of one Imprint with the second. This process is repeated for every possible combination of seizures for each subject resulting in a seizure-pair matrix (Fig. 1C) for DTW and OE-DTW, respectively. Note that while the DTW distance matrix is symmetric, the OE-DTW distance matrix is not symmetric as the algorithm only allows for one time series to be cut short. We defined a truncated-continuing pair as one in which the OE-DTW algorithm improved on the classical DTW algorithm by at least 20% (Fig. 1D).

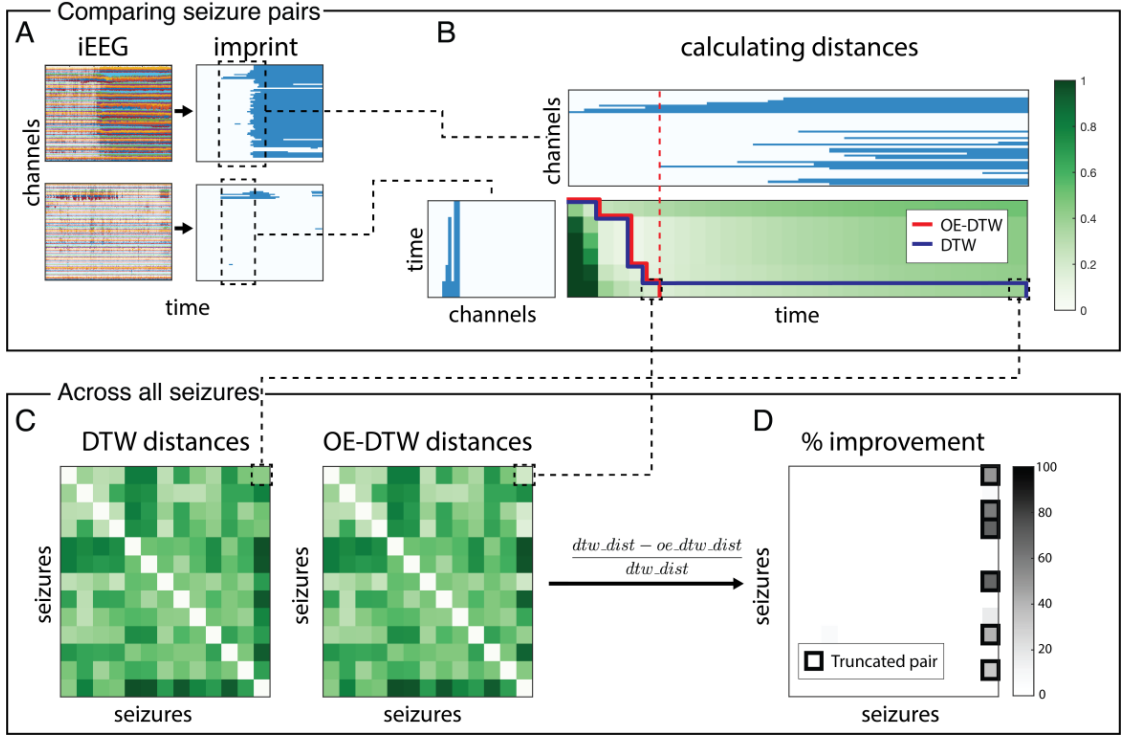

*Figure 1: Truncated-continuing seizure pair matching for an example subject.* A) Examples of continuing (top) and truncated (bottom) seizures from the same subject. Each iEEG signal (left) is pre-processed, then features are extracted and thresholded to generate a binary imprint (right). This imprint is then trimmed to the times of minimum and maximum spread (shown with black dotted lines). B) Simplified illustration of how the OE-DTW algorithm compares the trimmed imprints from panel A). For each time point in each imprint, the minimum cumulative distance from the start of both seizures is found. The OE-DTW algorithm (red) finds a minimum distance using every time point in imprint A, and any amount of imprint B. The proportion of imprint B used for this comparison is marked with a dotted red line. The DTW algorithm (blue) must include the whole of both imprints. C) This process is repeated, comparing every seizure recorded from this subject to every other. This produces a matrix of distances for both the DTW and OE-DTW algorithms. D) For each seizure pair the % improvement is calculated. If the % improvement is above the threshold of 20%, the seizure pair is labelled a truncated-continuing pair, shown with a black outline. For this example subject, the $13^{th}$ seizure was matched as a truncated seizure to the $1^{st}, 3^{rd}, 4^{th}, 7^{th}, 10^{th}$ and $12^{th}$ seizures, with the OE-DTW algorithm providing no improvement for the vast majority of the remaining seizure pairs.

## 2.6 Statistical analyses

### 2.6.1 Truncated-continuing pair occurrence

To test if ASM tapering during an iEEG implantation made pairs of truncated-continuing seizures more likely to occur, we compared the percentage of subjects that had truncated-continuing pairs detected in the tapered and non-tapered groups. A $\chi^2$ test was performed to determine if there was a significant difference between these two groups.

### 2.6.2 Relationship between ASM load and seizure duration

To compare seizure duration across subjects, we first took the natural log of each seizure duration, then normalised these values across subjects by taking the subject mean from each seizure duration.

Using the entire tapered cohort, we compared normalised log duration to ASM load using the following formula:

$$log\text{-}duration \sim \beta_0 + \beta_1 ASM\text{-}level \qquad (2)$$

We then repeated this analysis using only the seizures identified as part of any truncated-continuing pairs.

### 2.6.3 Differences between truncated and continuing seizures

First, we analysed the difference between seizure duration for truncated and continuing seizures. For each subject, we calculated the mean duration for all truncated seizures and continuing seizures separately. We then used a paired sample t-test to determine if there was a significant difference between the mean truncated duration and mean continuing duration across the cohort. Secondly, we repeated this analysis but instead calculated the mean ASM load for a subject's truncated seizures and continuing seizures for the identified continuing-truncated pairs. We then again used a paired sample t-test to determine if there was a significant difference between ASM load at the time of truncated seizures and continuing seizures across the cohort.

## 2.7 Code and data availability

Analysis code as well as anonymised data (Imprint and drug plasma concentrations) used are openly available on Github at https://github.com/NathanMEvans/Temporal-Modulation-truncation.

## 3. Results

### 3.1 Truncation detection is no more likely to occur in subjects undergoing ASM tapering

Of the 42 subjects with ASM tapering, 15 (36%) had truncated-continuing seizure pairs detected, compared to 7 (26%) of the 27 subjects that did not undergo tapering. There was no significant association between ASM tapering and truncated-continuing seizure pair occurrence ($\chi^2_{1,N=69}$ = 0.725, $p$ = 0.40), suggesting that factors other than ASM load affect the extent of seizure propagation.

## 3.2 Association between ASM load and duration persists for seizures in continuing-truncated pairs

We fitted the linear regression model described in equation 2 to all seizures in the tapered cohort (Fig. 2A), finding $\beta_1$ = -0.04 (p < 0.001). We then repeated this analysis on seizures that were part of any truncated-continuing seizure pair to test if this relationship persisted for this subsection of the data, finding a slightly larger correlation $\beta_1$ = -0.05 (p < 0.01).

## 3.3 Truncated seizures correlated with shorter seizure durations but not higher ASM loads

Across all truncated-continuing pairs, we calculated the mean duration for each subject's truncated and continuing seizures respectively (Fig. 2B). Continuing seizures were significantly longer ($t_{14}$ = 3.84, $p < 0.01$)

We calculated the mean ASM load at the time of each subject's truncated and continuing seizures (Fig. 2C). There was no significant difference between the truncated and continuing values ($t_{14}$ = -0.44, $p$ = 0.67).

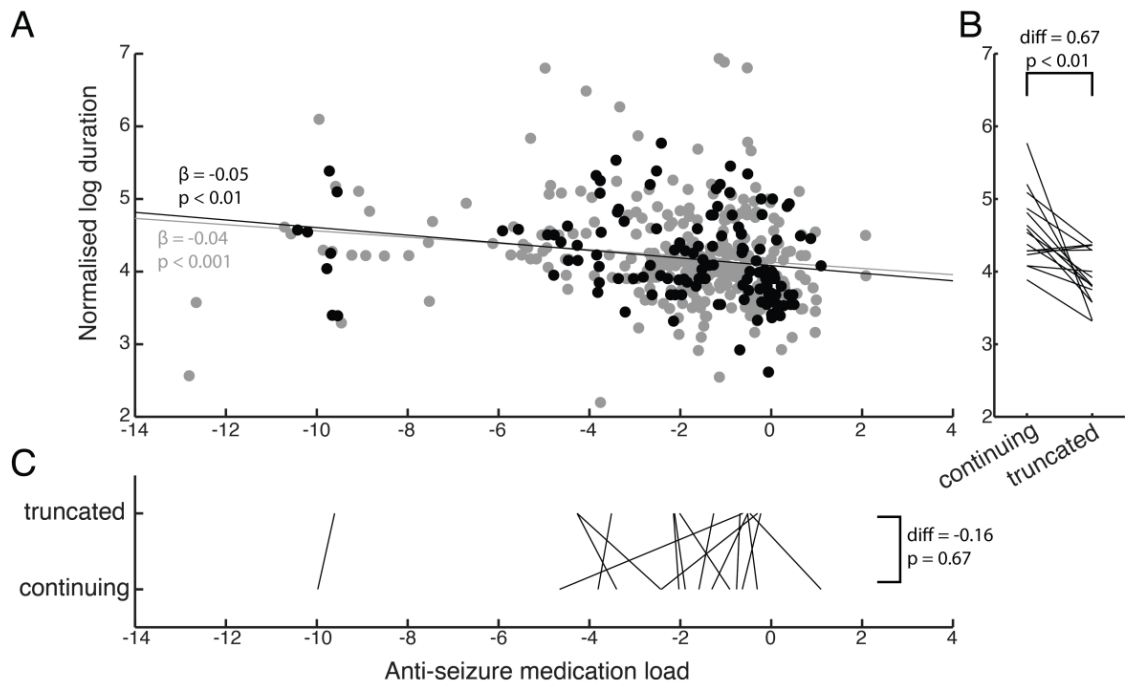

*Figure 2: Relationship between ASM load and duration for truncated and continuing seizure pairs.* A) ASM load plotted against log duration across all seizures (grey) and seizures involved in a truncated-continuing seizure pair (black). ASM load and log duration are significantly (p < 0.01) negatively correlated for both groups of seizures with $\beta_1$ value (see equation 2) of -0.04 and -0.05 respectively. B) Mean values for adjusted log duration for *continuing and truncated seizures for each subject. Each line connects the continuing and truncated value for a specific subject. Adjusted log duration was significantly (p < 0.01) higher for continuing seizures than truncated seizures. C) Mean values ASM load for continuing and truncated seizures for each subject. Each line connects the continuing and truncated value for a specific subject. There was no significant difference between ASM load for continuing seizures vs truncated seizures (p = 0.67).*

## 4. Discussion

We investigated the associations between ASM plasma concentration and seizure spread dynamics. We defined truncated-continuing seizure pairs based on a shared initial pathway, originally hypothesising that truncated seizures would occur at higher ASM loads than matched continuing seizures. However, subjects undergoing medication tapering were no more likely to have these truncated-continuing pairs than those without tapering, who had a much more constant medication load throughout the iEEG recording. Further evidence against our hypothesis is that, while the expected negative correlation between ASM load and duration exists in our dataset, and truncated seizures do tend to be shorter than their continuing counterparts, we found no significant difference between ASM loads at the time of truncated and continuing seizures. This suggests that the association between medication load and seizure severity is not driven by causing these pathway truncations, but by another mechanism.

Considering seizure spread patterns, shorter seizure durations could be a result of a seizure progressing faster while following the same spread patterns, or could be due to a seizure following a separate spread pattern that takes less time to progress from onset to termination. Different patterns could also be vary in the extent of their spread. Previous studies[16,17] found that medication withdrawal does not affect the onset localisation for surgery in most cases. However, in individuals in whom medication did affect onsets[18,19], a better understanding of how and when this occurs could be helpful when estimating the true epileptogenic zone.

Other factors could result in the occurrence of truncated seizures. Previous work[20] has shown that clinical seizure type can be partially predicted from the initial dynamics of the seizure onset. It may be possible to use a similar approach to predict when a seizure will truncate.

Increasing ASM doses is the current first-line approach for controlling seizures, but many people already take the highest tolerable dose. Therefore, other approaches, such as neuromodulation with vagus nerve stimulation may be helpful in some individuals[21]. It remains unclear why vegus nerve stimulation causes seizure reduction in some individuals, but the process likely involves blocking propagation in epileptic networks[2]. It is possible that neuromodulation promotes early termination of seizures in a similar manner to the seemingly naturally occurring truncated seizures observed in this study.

iEEG implantations give a unique insight into brain states and seizure dynamics which can be leveraged to understand many treatments. However, iEEG data obtained from those undergoing presurgical evaluation may not be representative of the wider epilepsy population. Less invasive modalities, such as scalp EEG and magnetoencephalography (MEG), may provide more generalisable insights into the relationship between ASMs and seizure severity.

A further challenge when analysing the effects of ASMs in this study, is disentangling their impact from other time-varying factors known to impact seizure dynamics, such as fatigue, anxiety and disturbance of the sleep-wake cycle[1,4]. These confounding factors make it difficult to isolate the effects of ASM tapering, especially when the duration of iEEG studies is relatively short. Future work could seek to separate these variables to better understand seizure dynamics during iEEG examinations.

In conclusion, more research into the effect of ASMs on seizure spread patterns is needed. Investigating truncated seizures recorded with iEEG may improve our understanding of the mechanisms modulating seizure termination, and aid in the conception of new therapeutic approaches.

## Acknowledgements

The author's contributions are given below.

- Conceptualisation: NE YW
- Methodology: NE SJG GMB CT
- Analysis: NE
- Data Acquisition: FAC BD JSD AWM AM JDT
- Data processing: NE SJ GMB GMS PNT YW
- Writing - Original Draft: NE
- Writing - Review & Editing: NE SJG GMS JSD YW
- Supervision: YW RT

We thank members of the Computational Neurology, Neuroscience & Psychiatry Lab (www.cnnp-lab.com) for discussions on the analysis and manuscript.

We further thank Charlotte McLaughlin and all the NHNN Telemetry Unit staff, and are grateful to all the people with epilepsy whose intracranial EEG data enabled this research.

P.N.T. and Y.W. are both supported by UKRI Future Leaders Fellowships (MR/Y034104/1, MR/V026569/1). JSD, JdT are supported by the NIHR UCLH/UCL Biomedical Research Centre. S.J.G is supported by the Engineering and Physical Sciences Research Council (EP/L015358/1) and ADLINK. N.E is supported by a Epilepsy Research Institute (ERI) Studentship.

## Disclosure of conflicts of interest statement

None of the authors have potential conflicts of interest to be disclosed.

## Ethical publications statement

We confirm that we have read the Journal's position on issues involved in ethical publication and affirm that this report is consistent with those guidelines.

## Figure Legends

**Figure 1: Truncated-continuing seizure pair matching for an example subject.** A) Examples of continuing (top) and truncated (bottom) seizures from the same subject. Each iEEG signal (left) is pre-processed, then features are extracted and thresholded to generate a binary imprint (right). This imprint is then trimmed to the times of minimum and maximum spread (shown with black dotted lines). B) Simplified illustration of how the OE-DTW algorithm compares the trimmed imprints from panel A). For each time point in each imprint, the minimum cumulative distance from the start of both seizures is found. The OE-DTW algorithm (red) finds a minimum distance using every time point in imprint A, and any amount of imprint B. The proportion of imprint B used for this comparison is marked with a dotted red line. The DTW algorithm (blue) must include the whole of both imprints. C) This process is repeated, comparing every seizure recorded from this subject to every other. This produces a matrix of distances for both the DTW and OE-DTW algorithms. D) For each seizure pair the % improvement is calculated. If the % improvement is above the threshold of 20%, the seizure pair is labelled a truncated-continuing pair, shown with a black outline. For this example subject, the $13^{th}$ seizure was matched as a truncated seizure to the $1^{st}, 3^{rd}, 4^{th}, 7^{th}, 10^{th}$ and $12^{th}$ seizures, with the OE-DTW algorithm providing no improvement for the vast majority of the remaining seizure pairs.

**Figure 2: Relationship between ASM load and duration for truncated and continuing seizure pairs.** A) ASM load plotted against log duration across all seizures (grey) and seizures involved in a truncated-continuing seizure pair (black). ASM load and log duration are significantly (p < 0.01) negatively correlated for both groups of seizures with $\beta_1$ value (see equation 2) of -0.04 and -0.05 respectively. B) Mean values for adjusted log duration for continuing and truncated seizures for each subject. Each line connects the continuing and truncated value for a specific subject. Adjusted log duration was significantly (p < 0.01) higher for continuing seizures than truncated seizures. C) Mean values ASM load for continuing and truncated seizures for each subject. Each line connects the continuing and truncated value for a specific subject. There was no significant difference between ASM load for continuing seizures vs truncated seizures (p = 0.67).